# CONTROL SYSTEM DESIGN PHILOSOPHY FOR EFFECTIVE OPERATIONS AND MAINTENANCE[*]

M. H. Bickley, K. S. White, Jefferson Laboratory, Newport News, VA 23606, USA

## Abstract

A well-designed control system facilitates the functions of machine operation, maintenance and development. In addition, the overall effectiveness of the control system can be greatly enhanced by providing reliable mechanisms for coordination and communication, ensuring that these functions work in concert. For good operability, the information presented to operators should be consistent, easy to understand and customizable. A maintainable system is segmented appropriately, allowing a broken element to be quickly identified and repaired while leaving the balance of the system available. In a research and development environment, the control system must meet the frequently changing requirements of a variety of customers. This means the system must be flexible enough to allow for ongoing modifications with minimal disruptions to operations. Beyond the hardware and software elements of the control system, appropriate workflow processes must be in place to maximize system uptime and allow people to work efficiently. Processes that provide automatic electronic communication ensure that information is not lost and reaches its destination in a timely fashion. This paper discusses how these control system design and quality issues have been applied at the Thomas Jefferson National Accelerator Facility.

## 1 INTRODUCTION

Jefferson Laboratory uses a control system based on the Experimental Physics and Industrial Control System (EPICS). At the lab, a number of different physical plants are controlled with EPICS, including an electron accelerator, a free electron laser, a helium liquification plant, and three experimental end stations[1]. In order to control the plants, the lab utilizes more than 50,000 I/O control points on 100 front-end computers. Operating these machines around the clock involves many operators and support personnel. As the number of hardware devices and users increases, the effort needed for coordination and communication can increase exponentially.

To combat these potential inefficiencies, Jefferson Lab has developed its control system with a philosophy that emphasizes the importance of enabling all users to work more effectively. This paper examines three broad categories of control system usage: operations, maintenance, and development. For each category, the paper will discuss administrative choices, support tools and aspects of control system design that contribute to improving user effectiveness and streamlining communications.

## 2 OPERATIONS

### 2.1 User Interfaces

The most basic interaction between users and the control system is through synoptic displays. The displays allow operators to monitor control system data points and modify machine parameters. When there is no consistent user interface design, the job of working with the control system is more difficult. At Jefferson Laboratory, several steps were taken to enable operators to work more effectively with MEDM, our synoptic display program. Interface developers have tools and an administrative framework that gives them the freedom to innovate, while ensuring that their products integrate well with the rest of the control system.

The first piece of the administrative framework was the development of interface standards. To ensure consistency, interface developers are given a palette of colors including obvious choices, for example red to indicate alarm conditions and yellow to indicate warnings. When screen items are user-modifiable their background is light blue, and if they are for display only they are dark blue. All interfaces include a color-coded title bar. This gives users a quick way to identify the system the screen controls.

To ensure that user interfaces support operations well, machine operators do a significant fraction of the interface development. They typically begin with screens provided by the device control developer. These screens are useful for managing a single device, or for use by experts, but are often too detailed to be used for operations. An operator modifies the low-level control screen and generates higher-level interfaces that, for example, show how the device integrates into the control system. An interactive GUI builder provides operators with the ability to easily develop interfaces optimized for their needs. The original expert screens remain available for debugging.

---

[*] This work was supported by the U.S. DOE Contact No DE-AC05-84-ER40150

As an additional aid to interface development, Jefferson Laboratory has created a library of object code that is used to generate synoptic display files programmatically. The library is useful with systems that include a large number of similar devices. For these systems, high-level displays can contain thousands of EPICS Process Variables. Rather than building screen files tediously by hand, one can easily write a program that generates a detailed screen using input from a data file. This is so simple that screen designers can invest the effort to create a very useful screen layout; confident their work will not have to be redone if the machine hardware is modified.

By using standards to ensure consistency, and providing tools for operators to customize screens for their needs, the task of interacting with the control system is simplified. Users can spend their energies understanding the meaning of the data, rather than deciphering its presentation.

## 2.2 Communications

Operating and maintaining Jefferson Lab's accelerator involves over 200 people in various groups. Coordination of such a large number of people can be problematic, as a great deal of information must be organized and made available to all staff members. Good communication is vital in using the control system to support smooth operation of such a large machine. To facilitate this, the laboratory has developed an electronic logbook (Elog) that is closely integrated with other control system tools. The Elog is used to ensure that all staff have easy access to information about the status of the accelerator.

An electronic logbook is superior to a paper version in several respects: It allows many people to generate and read entries in parallel. An electronic logbook can be updated via many methods, including automated generation of entries, web-based interfaces, and with customized log entry tools. The Jefferson Lab Elog can also be examined with many tools, including search engines, database forms, and a standard web browser.

For operations, the automatic log entry feature is an important capability. A significant fraction of the operators' time is spent recording information, to provide an accurate record of machine operations. Automated logging tools enable operators to efficiently make detailed entries with minimum distractions. These entries usually occur as a secondary output of a program that changes machine parameters or makes a measurement. This saves the operations staff time, since they do not have to make manual entries.

The most commonly used log entry tool is a custom interface named "dtlite". One of its features is the ability to capture X-windows screens. If an operator wishes to record the information presented in any window, he can associate the screen with a log entry. This enables him to show exactly what he sees at any time for later analysis. When encountering a problem, the operator can save the information quickly, with two mouse clicks, and move on to other tasks.

On occasion, while running the accelerator, the operations staff will identify a problem. They collect information about the problem, generate an Elog entry containing a text description and graphics if needed, and can, with dtlite, ensure that the problem will be addressed for maintenance. The same information that is placed in the Elog is added to a trouble-tracking database, called NEWTS. By selecting the appropriate button, the NEWTS entry can automatically record machine downtime associated with the reported problem. The operators do not need to duplicate any data entry efforts, because the Elog, NEWTS and downtime systems are integrated.

Jefferson Lab has used the Elog and other communication tools to help all control system users to be more effective. By ensuring that there is a consistent effort to address user communication issues during software design, the control system continuously enhances operational communication.

## 3 MAINTENANCE

Scheduled maintenance activities at Jefferson Laboratory include work to upgrade and enhance hardware or software, or to add or remove components. The lab has few scheduled maintenance opportunities, so it is vital to work efficiently. The result is that there is a strong motivation to do as much work as possible during a short period of time. It is also important to ensure that different maintenance tasks do not interfere with each other. For example, if one group is upgrading hardware in a portion of the machine, an unrelated software enhancement in the same area should not be attempted at the same time. A single person coordinates all accelerator maintenance activities to ensure that conflicts are minimized.

All support staff interested in performing repair, maintenance or upgrade work, submit their proposed activities to the maintenance coordinator. Control system tools are an integral part of this coordination effort, by enabling users to generate work plans that are automatically forwarded through the system.

Software maintenance is organized through the development of written test plans generated with web-based tools, which include several different templates. The templates provide button selections for common information and procedures for standard processes. The author lists the enhancements that will be achieved with the modification, what features it has, the expected test duration, and the steps required to install the upgrade and roll it back in the event that there are problems. The test plan is submitted, becomes viewable via the web, and an email notification is sent to a reviewer, who determines if plan is complete and reasonable, and either approves or returns it. If a test plan is returned, the reviewer indicates the deficiencies, and the author receives an email notification so that the plan can be improved. When approved, the test plan information is forwarded by email to the maintenance coordinator, and the work is scheduled. For each test plan completed during a maintenance period, an Elog entry is made that includes the test results The Elog then has a complete record of all of the maintenance activities.

As mentioned earlier, the Jefferson Laboratory control system spans a number of different physical plants. The control system is segmented so that each of the separate plants can function independently. The separate segments called "fiefdoms," make it possible to provide a consistent suite of tools while maintaining each plant's independence. Each fiefdom has all of the software and hardware required to function in isolation from the others. The segmentation is especially important during maintenance periods, because it enables portions of the control system to be unavailable in one fiefdom without impacting others. For example, if operating system patches must be installed, the modification can be made on one fiefdom at a time, and other fiefdoms can continue to operate, available for other maintenance work. During normal operation, all fiefdoms are accessible to all others. This makes it possible to centralize software development efforts in one fiefdom. When software is ready to be exercised operationally, it is easily distributed to the destination fiefdom. Support for the control system segmentation is built into the software development and operational tools used in the control system, making the segmentation transparent in normal operation.

## 4 DEVELOPMENT

A suite of tools is available to help software developers work with the control system. A primary goal of any operational control system is to enable developers to easily install software for testing and roll back these changes. At Jefferson Lab, where testing facilities are limited, the operational machine is usually the integration test bed for software enhancements. To minimize the impact of testing on availability, the lab uses a well-designed version management system for its low-level applications[2].

The purpose of the tools is to assist the developers in organizing and managing their applications while maintaining flexibility. A well-defined framework enables programmers to have a great deal of flexibility with the application implementation details. This is important, because the scope of various applications can be so different, ranging from an application that controls a single device with a few I/O channels, to one that drives an RF system with thousands of control points. A well-designed structure enables all to coexist, and aids in support efforts. The organizational similarities between applications provide a basis of understanding for all developers. This makes it easier for on-call staff to support a large body of applications. Also, because of the commonalities, a new developer can more quickly come up to speed.

The tools support versioning of applications, and the associated installation and rollback needed for support of the operational software. The computer scientist can create new application versions as needed, and designate which versions are operationally valid, using the test plan tools described above. The list of valid versions is stored in a database, accessible through web-based tools. Any software developer, can, with information provided by the database and using the standard tools, select a different software version.

Each step of the software development process is documented, with support of the available tools. This documentation is automatically available via the web, so that any control system user can find design and release notes for any version of the software. The easy accessibility helps users to understand the capabilities of the control system software.

## CONCLUSIONS

A control system that is highly effective does more than simply provide device control and user interfaces for operating a complex, automated machine. A good control system enhances the productivity and quality of the work of all users, including operators, maintainers and developers. It provides mechanisms for supporting communication and coordination between people. This enables users to focus on the task at hand, rather than spending their energies in bookkeeping and other important but mundane tasks. The effective control system provides consistency in the appearance and behavior of tools for users, helping them to work more efficiently. Finally, it can also facilitate the work of enhancing and maintaining the accelerator by simplifying the work of technical support staff and operators.

## REFERENCES

[1] K.S. White, M. H. Bickley, "Control System Segmentation", PAC 01, Chicago, USA, June 2001.

[2] S. Schaffner, M. Bickley, et. al., "Tools for Application Management at Jefferson Lab", ICALEPCS 99, Trieste, Italy, October 1999.